%% file: paper-LHY-crossover.tex
\documentclass[floatfix,aps,twocolumn,amsmath,amssymb,superscriptaddress,prl]{revtex4-1}
\usepackage{bm,bbm}
\usepackage{amssymb,amsfonts,amsmath}
\usepackage{graphicx}
\usepackage{xcolor}
\usepackage{ulem}
\usepackage{SIunits}
\usepackage[utf8]{inputenc}
\usepackage[english]{babel}
\usepackage{hyperref}
\usepackage{ulem}
\usepackage{braket}
\usepackage{bpchem}
\graphicspath{ {./Bilder/} }
\usepackage{tikz}
\usepackage{pdfpages}
\usepackage{url}

\makeatletter
\AtBeginDocument{\let\LS@rot\@undefined}
\makeatother

\newcommand{\rs}[1]{\rm{\scriptscriptstyle #1}}

\newcommand{\dd}{d}

\newcommand{\RM}[1]{\MakeUppercase{\romannumeral #1}}
\newcommand{\slt}{a_{s}}

\newcommand{\lp}{l_\perp}

\newcommand{\sump}{\sideset{}{'}\sum}

\begin{document}

\title{Dimensional crossover for the beyond-mean-field correction in Bose gases
}
\author{Tobias Ilg}
\affiliation{
Institute for Theoretical Physics \RM3 and Center for Integrated Quantum Science and Technology, University of Stuttgart, DE-70550 Stuttgart, Germany}

\author{Jan Kumlin}
\affiliation{
Institute for Theoretical Physics \RM3 and Center for Integrated Quantum Science and Technology, University of Stuttgart, DE-70550 Stuttgart, Germany}

\author{Luis Santos}
\affiliation{
Institut f\"ur Theoretische Physik, Leibniz Universit\"at Hannover, Appelstr. 2, DE-30167 Hannover, Germany}

\author{ Dmitry S.~Petrov}
\affiliation{
LPTMS, CNRS, Univ. Paris Sud, Universit\'e Paris-Saclay, 91405 Orsay, France}

\author{Hans Peter B\"uchler}
\affiliation{
Institute for Theoretical Physics \RM3 and Center for Integrated Quantum Science and Technology, University of Stuttgart, DE-70550 Stuttgart, Germany}

\date{\today}

\begin{abstract}
We  present a detailed beyond-mean-field analysis of a weakly interacting Bose gas in the crossover from three to low dimensions. We find an analytical solution for the energy and provide a clear qualitative picture of the crossover in the case of a box potential with periodic boundary conditions. We show that the leading contribution of the confinement-induced resonance is of beyond-mean-field order and calculate the leading corrections in the three- and low-dimensional limits. We also characterize the crossover for harmonic potentials in a model system with particularly chosen short- and long-range interactions and show the limitations of the local-density approximation. Our analysis is applicable to Bose-Bose mixtures and gives a starting point for developing the beyond-mean-field theory in inhomogeneous systems with long-range interactions such as dipolar particles or Rydberg-dressed atoms.

\end{abstract}

\maketitle
{\it Introduction.} Recent experiments have demonstrated the stabilization of quantum droplets with weakly interacting gases with dipolar interactions \cite{Kadau2016,Schmitt2016,Ferrier-Barbut2016,Chomaz2016}, as well as Bose-Bose mixtures \cite{Cabrera2017,Cheiney2018,Semeghini2018}.
The basic ingredient  for this remarkable phenomenon is the near cancellation of the mean-field interaction such that the beyond-mean-field corrections become relevant, prevent a collapse of the Bose gas, and eventually lead to the formation of droplets \cite{Petrov2015}. 
These quantum droplets have renewed the interest in understanding beyond-mean-field effects \cite{Petrov2015,Edler2017,Baillie2017}, but
although these droplets are inhomogeneous and anisotropic, 
the theoretical analysis so far has mostly been based on the local-density approximation.
In this Rapid Communication, we present the behavior of the beyond-mean-field correction beyond local-density approximation for a dimensional crossover.

The leading correction to the ground-state energy for a weakly interacting Bose gas in three dimensions has been pioneered by Lee, 
Huang, and Yang (LHY) \cite{Lee1957,Lee1957a},  
and has been well confirmed experimentally with cold atomic gases \cite{Navon2011,Lopes2017}, and the analysis has been extended to dipolar interactions \cite{Lima2012}. 
For lower dimensions, its behavior is well understood from the exactly solvable theory by Lieb  and Liniger \cite{Lieb1963,Lieb1963a} in one dimension, while the behavior in
two dimensions has been  studied in detail in the past \cite{Schick1971,Popov1972,Cherny2001,Mora2003,Pricoupenko2004,Astrakharchik2009,Mora2009}.  On a theoretical level, the difficulties appear by the proper  renormalization of the contact interaction.  
A natural approach to avoid this problem is the application of the theory developed by Hugenholtz and Pines \cite{Hugenholtz1959}.
This approach is highly suitable for the study of the crossover and has recently been applied for dipolar systems for a qualitative analysis of the crossover \cite{Edler2017}.

\begin{figure}
   \includegraphics[scale=0.56]{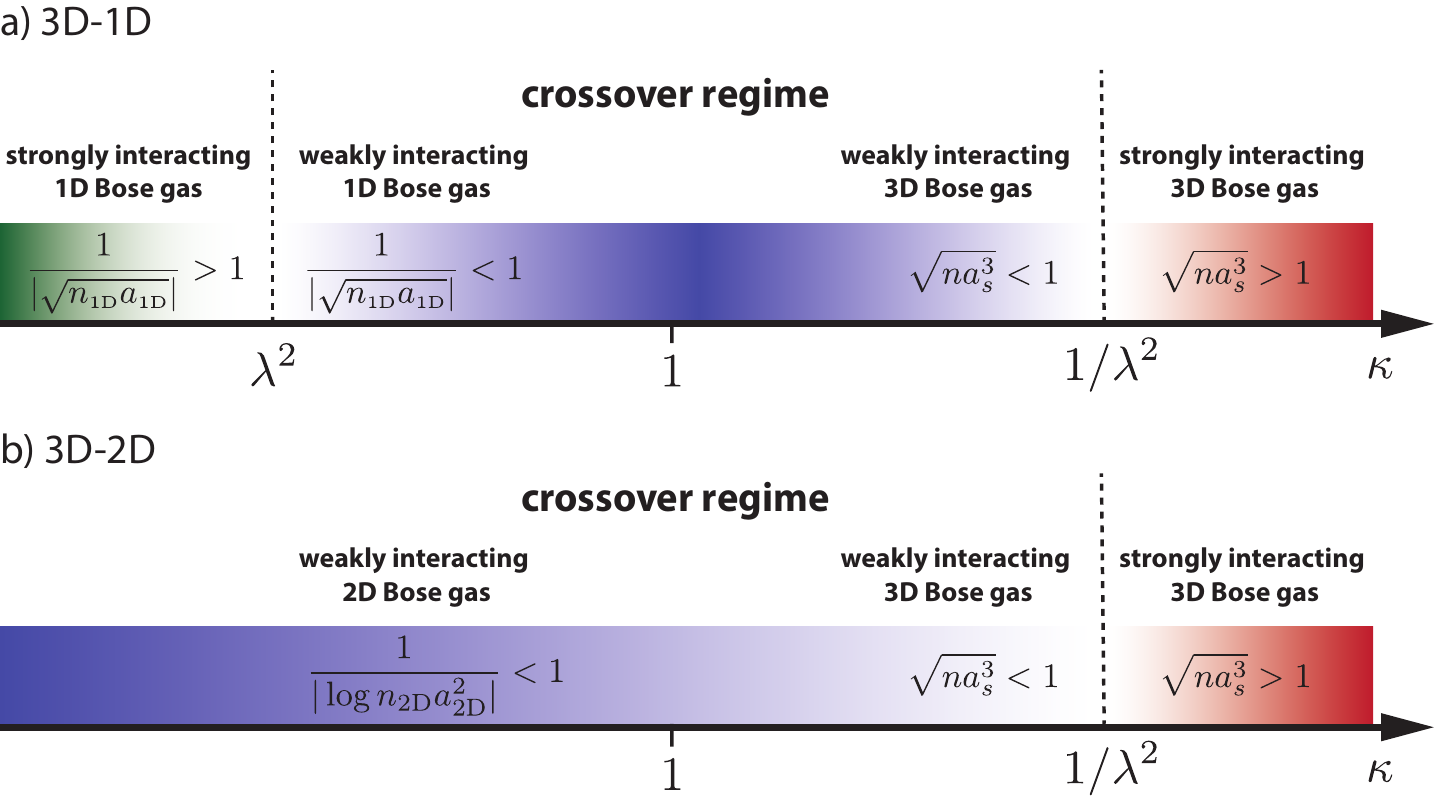}
\caption{Crossover regime from three dimensions to a) one dimension and b) two dimensions: the parameter characterizing the crossovers
is $\kappa=n\slt\lp^2$, which relates the transverse confinement to the chemical potential. The beyond-mean-field predictions 
are valid in the weakly interacting regime (blue), which requires that $\sqrt{n a_{s}^3}  \ll 1$ for three dimensions, 
$1/| n_{\rs{1D}} a_{\rs{1D}}|\ll  1$ for one dimension, and $1/|\ln n_{\rs{2D}} a^2_{\rs{2D}}| \ll 1$ for two dimensions. These requirements translate to the conditions $\lambda^2 \ll \kappa \ll 1/\lambda^2$ for the 3D-1D crossover and $\kappa\ll1/\lambda^2$ for the 3D-2D crossover, where $\lambda=\slt/\lp$.
Furthermore, it shows that $\lambda$ is our small parameter and we require $\lambda \ll 1$ for the validity of our results.
 \label{fig1}}
\end{figure}

In this Rapid Communication, we calculate the beyond-mean-field correction for a one-component weakly interacting Bose gas with scattering length $a_{s}$ in the crossover from three to two and from three to one dimension. We start with a system that  is confined along one or two directions by a box potential with length $l_\perp$ and periodic boundary conditions. In this case the ground state is homogeneous and is characterized by the density $n$. 
We calculate the LHY correction as a function of the dimensionless parameters $\lambda=a_{s}/l_\perp$ and $\kappa=na_{s}l_\perp^2$ (see Fig.~\ref{fig1}). 
The result is expressed in terms of known special functions allowing for an analytical description of both limits. In particular, on the low-dimensional side of the crossover (small $\kappa$) this method reproduces results obtained for purely low-dimensional models with two-body interactions correctly renormalized by the confinement \cite{Olshanii1998,Petrov2001}. 
We extend this crossover analysis to the case of a harmonic confinement by using a particular form of the interaction potential, which guarantees the Gaussian shape of the mean-field ground-state wave function along the crossover, and derive the LHY corrections beyond local-density approximation.

The small parameter ensuring the weakly interacting regime for a three-dimensional Bose gas
is $\sqrt{n a_{s}^3} \ll 1$. This condition can either be satisfied by small scattering lengths
or in the dilute regime. With decreasing the density in the presence of a transverse confinement, 
the system crosses over into the low-dimensional regime  
signaled by the condition that the transverse level spacing $E_{\perp}= \hbar^2 /m l_\perp^{2}$ is 
comparable to the chemical potential $\mu =4 \pi \hbar^2 a_{s} n /m$ giving rise to a dimensionless parameter
$\kappa = n a_{s} l_{\perp}^{2} \sim \mu /E_{\perp}$: The regime  $\kappa \gg 1$ describes a 
three-dimensional setup, while for $\kappa \ll 1$ the system is effectively lower dimensional (see Fig.~\ref{fig1}).  
Furthermore in one dimension, it is well established that the strongly correlated Tonks-Girardeau regime is reached in
the low-density limit, while weakly interacting bosons appear for high densities $|n_{\rs{1D}} a_{\rs {1D}}| \gg 1$; 
here, $n_{\rs{1D}}=n l_{\perp}^2$ is the one-dimensional density and the 1D scattering length $a_{\rs{1D}}=- l_{\perp}^2/2 \pi a_{s}$. Therefore, 
the proper small parameter controlling the weakly interacting regime in the full crossover regime is given by 
$\lambda= a_{s}/l_{\perp} \ll 1$ , and $\kappa$ is restricted to stay in the interval  $\lambda^2 \ll \kappa\ll 1/\lambda^2$ (see Fig.~\ref{fig1}).
By contrast, there is no lower-density limit for the weakly  interacting regime in two dimensions and one requires
 $\kappa\ll1/\lambda^2$.
Written in terms of $\lambda$ and $\kappa$, the beyond-mean-field correction in the three-dimensional regime takes the form \cite{Lee1957,Lee1957a}
\begin{equation}
  E_{\rs{3D}}=\frac{2\pi\hbar^2}{m}\frac{V}{\lp^4 \slt}
	\left( \kappa^2+\lambda \: \frac{128}{15\sqrt{\pi}}\:\kappa^{5/2} 
	\right),
	\label{eq:threedimenisonalenergy}
\end{equation}
where $V$ denotes the volume of the system.
In the following, it is convenient to express the ground-state energy $E[\kappa,\lambda]$ in the crossover
in terms of the energy scale  $E_{0}=\frac{2\pi\hbar^2}{m}\frac{V}{\lp^4 \slt} $.

Our approach for the determination of the beyond-mean-field correction in the crossover is based on the theory developed 
by Hugenholtz and Pines \cite{Hugenholtz1959}.
This method is equivalent to the  conventional approach 
based on the Bogoliubov theory \cite{Landaubook9}, but naturally avoids difficulties with ultraviolet divergencies. One can understand
this behavior as the divergencies in the Bogoliubov theory only provide a correction to the mean-field term 
proportional to $n^2$. 
In the approach of Hugenholtz and Pines the ground-state energy is determined by a differential equation, that does not determine this mean-field term.
From the differential equation and fixing the correct mean-field behavior in three dimensions,
we obtain the ground-state energy (see Supplemental Material)
\begin{equation} 
  E=   E_{\rs{3D}}+\kappa^2 \lambda  \:  E_{0} 
	\int_{\kappa}^{\infty}\!\!\!\! \dd\kappa'\, \left[ h(\kappa')-h_{\rs{3D}}(\kappa') \right].
	\label{eq:generalsolutionkappastar1}
\end{equation}
The first term in the integral  accounts for the correction to the beyond-mean-field contribution in the crossover
and takes the form 
\begin{equation}
    h(\kappa) = \frac{1}{\kappa^3 }\sum \hspace{-13pt}  \int   \left[\frac{2\varepsilon^2+3\kappa \varepsilon}{\sqrt{\varepsilon^2+2\kappa \varepsilon}}-2 \varepsilon-\kappa\right].
    \label{hvalues}
 \end{equation}
Here,  $\varepsilon= \pi\left( u^2+v^2+w^2 \right)/2$ is the single-particle excitation spectrum for periodic boundary conditions
 with $u$, $v$, and $w$ the three components of the single-particle 
momentum in dimensionless units. Furthermore, the notation $\sum \hspace{-10pt} \int  $ describes 
a summation over the transversal confined degrees of freedom, and 
an integration over the unconfined dimensions.
The last term in the integral in Eq.~(\ref{eq:generalsolutionkappastar1}) guarantees that the integral is convergent and vanishes for $\kappa \rightarrow  \infty$. It 
takes the form $h_{\rs{3D}}(\kappa) = -\frac{64}{15\sqrt{\pi \kappa}} $, and is the corresponding 
expression  to $h(\kappa)$ for a three-dimensional system.

\begin{figure*}[t]
\includegraphics[width= 1 \textwidth]{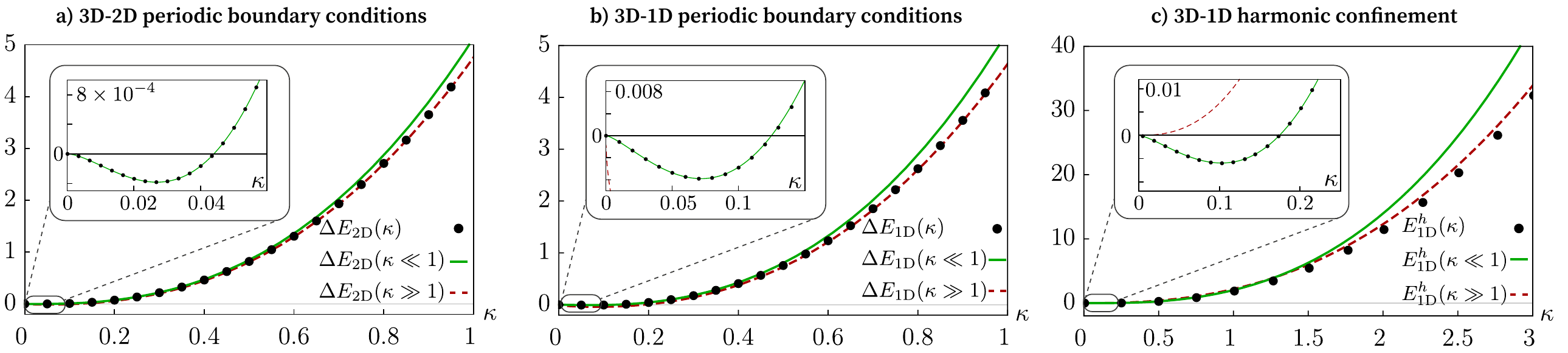}
\caption{Crossover behavior of the beyond-mean-field correction. Black dots denote the result from the numerical evaluation within the crossover, while 
the asymptotic behavior for small $\kappa$ is plotted as a green (solid) line,  and the red  (dashed) line shows the analytical prediction for large $\kappa$. (a) The behavior 
of the crossover from 3D to 2D with periodic boundary conditions in transverse direction.  (b) The behavior 
of the crossover from 3D to 1D with periodic boundary conditions in transverse direction.   (c) The behavior 
of the crossover from 3D to 1D with a harmonic trapping potential in transverse direction. Note, that $E_{\rs{1D}}^{h}(\kappa \ll 1)$ includes in addition to the analytical expression Eq.~(\ref{eq:harmonic1D}), the term $B_{\rs{1D}}^{h} \kappa^3$.
 \label{fig2}}
\end{figure*}

{\it Crossover to two dimensions.} We start with the dimensional crossover from a
three-dimensional setup toward a two-dimensional slab with just a single transverse dimension confined. 
The ground-state energy in the crossover is denoted as  $E_{\rs{2D}}$. 
Then,  $\sum \hspace{-10pt} \int =  \sum_{w} \int du\,dv$ and the integrations  $\int du\,dv$ can be performed 
analytically leading to
\begin{equation}
h(\kappa)=  \frac{1}{\kappa^3} \left[g(0)+ 2 \sum_{w=1}^{\infty} g\left(\frac{\pi w^2}{2}\right)  \right]
\end{equation}
with $g(q^2)=  2 q^2 \left( q^2  \!+\! \kappa \!-\! q \sqrt{q^2 \!+\!2 \kappa}\right)\! -\! \kappa^2$.
The summation can be performed by introducing a contour integral  $\sum_{w=1}^{\infty}   = -\frac{i}{2}\oint_{\gamma} dw \,\cot(\pi w) $  with the contour $\gamma$
surrounding the positive real axis. Then, only the integration along the branch cut of $\sqrt{q^2+2 \kappa}$ contributes, while the pole at $w=0$ cancels $g(0)$
in the summation. By this procedure, we obtain
\begin{equation}
  h(\kappa) = - \frac{32}{\sqrt{ \pi \kappa}}  \int_{0}^{1} dq \; q^3 \sqrt{1-q^2} \coth(q \sqrt{4 \pi \kappa}).
\end{equation}
As a consequence,  the ground-state energy $E_{\rs{2D}}$ in the crossover from three to two dimensions takes the form
\begin{align}
  \frac{E_{\rs{2D}}}{E_{0}}=&\kappa^2+\lambda\frac{128}{15\sqrt{\pi}}\kappa^{5/2}\nonumber
  \\
  &+  \lambda \kappa^2 \frac{32 }{\pi }   \int_{0}^{1} \hspace{-5pt}dq   \; q^2 \sqrt{1-q^2} \: \ln \left(1-e^{-\sqrt{16 \pi  \kappa}\: q } \right) \nonumber
\end{align}
and is shown in Fig.~\ref{fig2}; note that it is most convenient to show only the beyond-mean-field 
corrections $\Delta  E_{\rs{2D}}= (E_{\rs{2D}} /E_{0}-\kappa^2)/\lambda$.
In the three-dimensional regime with $\kappa\gg1$,  the transverse confinement leads to an attractive 
correction to the ground-state  energy, 
\begin{equation}
  \frac{E_{\rs{2D}}-E_{\rs{3D}}}{E_{0}}\overset{\kappa\gg 1}{=}  - \frac{\pi^{3/2}}{90} \: \lambda \:  \sqrt{\kappa} + O(
  \lambda \kappa^{-1/2}),
  \label{eq:groundstateenergy3d2d}
\end{equation}
which is nonvanishing even for $\kappa \rightarrow \infty$.  
For the quasi-two-dimensional Bose gas with $\kappa \ll 1$, the ground-state energy behaves as 
\begin{equation}
  \frac{E_{\rs{2D}}(\kappa)}{E_{0}}\overset{\kappa\ll1}{=}  \kappa^2+  \lambda  \kappa^2 \ln(\kappa4\pi\sqrt{e})  +  \frac{2 \pi}{3} \; \lambda \; \kappa^3  + O(\lambda \kappa^4).
  \label{eq:groundstateenergy2d}
\end{equation}
This expression accounts for the negative  beyond-mean-field correction to the ground-state energy as well 
as the zero crossing [see Fig.~\ref{fig2}(a)]. The third term describes an effective  three-body interaction due to quantum fluctuations, while the first two terms provide the ground-state energy of a purely two-dimensional Bose gas.
In order to establish the latter connection, we note that the ground-state energy  of a two-dimensional Bose gas 
takes the form \cite{Cherny2001,Mora2003,Pricoupenko2004,Mora2009,Astrakharchik2009} 
\begin{equation}
  \frac{E}{L^2}=\frac{2\pi\hbar^2n_{\rs{2D}}^2/m}{\ln(\frac{1}{n_{\rs{2D}}a_{\rs{2D}}^2})+
  \ln\big[\ln(\frac{1}{n_{\rs{2D}}a_{\rs{2D}}^2})\big]-\ln(e^{2 \gamma} \pi \sqrt{e})}
  \label{eq:twodimgroundstate}
\end{equation}
with the two-dimensional density $n_{\rs{2D}} = n l_{\perp}$ and the two-dimensional scattering length $a_{\rs{2D}}$.  
The connection between the $s$-wave scattering length $a_{s}$ and $a_{\rs{2D}}$ in confined systems has been derived for a harmonic trapping potential \cite{Petrov2001,Pricoupenko2008}; its generalization to periodic boundary conditions is straightforward $a_{\rs{2D}}= 2 l_{\perp} e^{-l_{\perp}/ 2a_{s}} e^{-\gamma}$ 
(see Supplemental Material) and was first obtained in \cite{Lammers2016}.
Inserting this $a_{\rs{2D}}$ into the ground-state energy of the two-dimensional Bose gas Eq.~(\ref{eq:twodimgroundstate})
and performing an expansion in the small parameter $\lambda=a_{s}/l_{\perp}$, we reproduce the first two terms in 
Eq.~(\ref{eq:groundstateenergy2d}). Note that the expansion (\ref{eq:groundstateenergy2d}) implies $\kappa \gtrsim \lambda e^{-1/\lambda} $.
Otherwise, for exponentially low densities, the two-dimensional small parameter   $1/|\ln n_{\rs{2D}} a^2_{\rs{2D}}| = 1/|\ln(\kappa/\lambda)- 1/\lambda| \ll 1$ \cite{Schick1971} is no longer dominated by $\lambda$ but rather by the logarithm of the densities.

{\it Crossover to one dimension.} 
Next we focus on the dimensional crossover from a three-dimensional setup toward a one-dimensional tube with two  transverse dimensions confined. 
Like the Bogoliubov theory, the method used here relies on the existence of a condensate, which is absent in one dimension.  
  In the weakly interacting regime, however, the ground-state energy is well described within the Bogoliubov theory, as the system still exhibits quasi-long-range order \cite{Petrov2000}.
The ground-state energy in the crossover is denoted as  $E_{\rs{1D}}$, and the evaluation of
$h(\kappa)$ requires $\sum \hspace{-10pt} \int =  \sum_{w,v} \int du$. 
In contrast to the 2D situation, we first perform the integration over $\kappa'$. Then, the expression for the beyond-mean-field correction reduces to 
\begin{equation}
  \frac{E_{\rs{1D}}-E_{\rs{3D}}}{E_{0}} =  \lambda \int du \left[  \sum_{v,w} f(\epsilon) - \int dv  dw 
  \:  f(\epsilon) \right],  \nonumber
\end{equation}
where $f(\epsilon) = \sqrt{\epsilon^2 + 2 \kappa \epsilon} - \epsilon - \kappa$.
Again, it is possible to derive an expression in well-known functions by performing the double sum
(see Supplemental Material)
\begin{align}
   \frac{E_{\rs{1D}}}{E_{0}}  =&  \kappa^2 (1+\lambda C_{\rs{1D}})- \lambda\frac{8}{3\sqrt{\pi}}\kappa^{3/2}  +\lambda \kappa^{2} \int_{0}^{\infty}\frac{ d\tau }{\sqrt{ \pi \tau}} \label{1Dresult}\\ &\times [ \vartheta_{3} \left(0, e^{-\tau}\right)^2\!-\!1]  \left[ 1\!- \!\frac{ e^{-2 \tau \kappa/\pi}I_{1}\left(\frac{2 \tau \kappa}{\pi}\right)}{\tau \kappa /\pi}\right]  \nonumber.
\end{align}
Here, $\vartheta_{3}(z,q)=\sum_{n} q^{n^2} \cos (2 n z)$ denotes the Jacobi theta function, while $I_\nu(z)$ is the modified Bessel function. 
The term involving $C_{\rs{1D}}$ is a shift to the mean-field energy due to the confinement defined as
\begin{equation}
  C_{\rs{1D}}=\int dv dw \frac{1}{\sqrt{v^2+w^2}}
  -\sump_{v,w}\frac{1}{\sqrt{v^2+w^2}}\\
  \approx3.899, \nonumber
\end{equation}
where the summation {\footnotesize $\sump$} omits the term $v\!=\!w\!=\!0$. 
The beyond-mean-field correction  $\Delta  E_{\rs{1D}}= (E_{\rs{1D}} /E_{0}-\kappa^2)/\lambda$ 
along the crossover is shown in  Fig.~\ref{fig2}(b).

Using  the properties of
Bessel functions, we expand Eq.~(\ref{1Dresult}) for small values of $\kappa \ll 1$ and we obtain the leading corrections
for  the one-dimensional regime
\begin{equation}
  \frac{E_{\rs{1D}}}{E_{0}}\overset{\kappa\ll1}{=}\kappa^2\left( 1+\lambda C_{\rs{1D}} \right)-\lambda\frac{8}{3\sqrt{\pi}}\kappa^{3/2}  + \lambda 
 B_{\rs{1D}}\kappa^3 + O(\lambda \kappa^4)
  \label{eq:energyonedim}
\end{equation}
with $B_{\rs{1D}}=(1/\pi) \text{\footnotesize{\(\sump\)}}_{vw} (v^2+w^2)^{-3/2} \approx 2.88$.  
These terms account for the attractive part of the beyond-mean-field correction as well as the zero crossing [see Fig.~\ref{fig2}(b)].
Again, we have a very clear interpretation of these results:  the term with $\kappa^3$ provides  an effective three-body interaction, while the other terms account for the ground-state
energy of a one-dimensional Bose gas.  The latter is well established to take the form \cite{Lieb1963}  
\begin{equation}
  \frac{E}{L}=-\frac{\hbar^2}{m a_{\rs{1D}}} n_{\rs{1D}}^2\left( 1-\frac{4\sqrt{2}}{3\pi}\frac{1}{\sqrt{|n_{\rs{1D}} a_{\rs{1D}}}|} \right).
  \label{eq:groundstatenergy1d}
\end{equation}
Including the effect of the confinement-induced resonance \cite{Olshanii1998}, 
the relation between $a_{\rs{1D}}$ and the $s$-wave scattering length $a_{s}$
takes the form $a_{\rs{1D}}=-\frac{1}{2\pi}\frac{\lp^2}{\slt}\left( 1-C_{\rs{1D}}\frac{\slt}{\lp} \right)
$ (see Supplemental Material).
Note, that the parameter $C_{\rs{1D}}$ is modified for periodic boundary conditions compared to a harmonic trapping \cite{Olshanii1998}.
Expanding the ground-state energy of a 1D Bose gas in the small parameter $\lambda= a_{s}/l_{\perp}$ therefore provides the first two terms in Eq.~(\ref{eq:energyonedim}), i.e.,
it naturally includes the leading contribution of the confinement-induced
resonance.
 Finally, it is also possible to provide an analytical expansion in the three-dimensional (3D) regime for $\kappa \gg 1$, and the leading correction is attractive:
\begin{equation}
  \frac{E_{\rs{1D}}-E_{\rs{3D}}}{E_{0}} \overset{\kappa\gg1}{=} - \sqrt{\kappa } \lambda  \frac{A_{\rs{1D}}}{2 \pi^{5/2}}  + O(\kappa^{-1/2})
  \label{eq:groundstateenergy3d1d}
\end{equation}
with the constant $A_{\rs{1D}}= \text{\footnotesize{\(\sump\)}}_{v,w} \left(v^2+w^2\right)^{-2} \approx 6.0268$.

{\it Harmonic confinement.} 
Finally, we apply our  understanding of the crossover to a system with harmonic confinement in the transverse directions; 
the trapping frequency is related to the transverse confinement  length via  $\omega_{\perp}= \hbar/ m l_{\perp}^2$.
We are not interested in mean-field modifications of the ground-state wave function due to interactions, 
which  was studied previously \cite{Muryshev2002,Sinha2006,Mazets2008}, 
but rather on beyond-mean-field corrections for a setup,
where the condensate remains in the lowest state of the harmonic confinement. 
Experimentally, this goal can be achieved for bosonic mixtures \cite{Petrov2015}.
On the theoretical level, this goal is conveniently achieved by adding an attractive interaction potential, which is dominated by a large range $r_{0} \gg l_{\perp},l_{\perp}/\sqrt{\kappa}
$  within the tube elongated along $x$, e.g., $ -4 \pi \hbar^2 a_{s} /(r_{0}\sqrt{\pi}m)\delta(y)\delta(z) e^{- x^2/r_{0}^2} $. 
Note, that such a potential does not contribute to the beyond-mean-field corrections due to its large range, but guarantees 
that the condensate remains in the lowest energy state of the transverse confinement within mean-field theory.

We start with the analysis of the dimensional crossover from three to one dimension with harmonic confinement. Then,  it is convenient to define
$\kappa = n_{\rs{1D}} a_{s}$,  while the single-particle excitation spectrum in dimensionless units is modified to 
$\epsilon = \pi /2 \left[ u^2 \!+ \!\left( v \!+\! w\right)/2\pi^2\right]$ with $v,w\in\{0,1,2,\ldots\}$, i.e., $v$ and $w$ denote  the quantum numbers for 
the harmonic oscillator modes of the transverse confinement.  The interaction potential will lead to mixing of
different transverse modes, and therefore, the Bogoliubov transformation in general involves many transverse modes, i.e., 
$a_{u,\alpha \beta}=\sum_{vw} \left[u_{u,vw}^{\alpha\beta}b_{u,vw}- v_{u,vw}^{\alpha\beta}b^{\dag}_{-u,vw}\right]$ with $a_{u, \alpha \beta}$ the new bosonic operators, 
and $u_{u,vw}^{\alpha\beta}$  ($v_{u,vw}^{\alpha\beta}$) the factors of the Bogoliubov transformation. 
Then, the term $h(\kappa)$ within the approach by Hugenholtz and Pines takes the form (see Supplemental Material)

\begin{equation}
 h(\kappa) =  \frac{4 \pi}{\kappa^3} \int du \sum_{\alpha \beta} \sum_{v w} \left[\epsilon - E_{\alpha \beta}\right] |v_{u,vw}^{\alpha \beta}|^2
 \label{hharmonic}
\end{equation}
with $E_{\alpha \beta}$ the Bogoliubov excitation energy in dimensionless units.  In general, the determination of the Bogoliubov excitation 
spectrum $E_{\alpha\beta}$ and the factors $v_{u,vw}^{\alpha \beta}$ requires a numerical analysis. 
The beyond-mean-field correction to the ground-state energy is determined by fixing the correct mean-field term $\propto \kappa^2$. In the previous analysis, we observed that the beyond-mean-field correction includes the modification of 
 the mean-field  term by the confinement-induced resonance. This allows us to fix the mean-field term in the one-dimensional regime $\kappa \ll 1$.
 Alternatively, one would expect that for a very shallow trapping potential the local-density approximation is well justified, which in turn allows us to
 fix the mean-field term for $\kappa \gg 1$; our numerical analysis shows that both approaches coincide. The result of this numerical
analysis in the full crossover from three to one dimension is shown in Fig.~\ref{fig2}(c). In the three-dimensional regime with $\kappa \gg 1$, 
we find excellent agreement between the numerical analysis and the prediction within the local-density approach. The latter is obtained by
integrating the 3D LHY result in Eq.~\eqref{eq:threedimenisonalenergy} over the transversal density profile  of the condensate, i.e.,
$  E_{\rs{1D}}^{h} =  \lambda \frac{512}{75\pi }\, \kappa^{5/2} E_{0}^{h}$  for $\kappa \gg 1$ with $E_{0}^{h}= \hbar \omega_{\perp} L/a_{s}$.
Note, that the term $\kappa^2$ is missing due to our special choice of the interaction potential.
In turn, for $\kappa \ll 1$ it is possible to derive the leading
corrections to the ground-state energy by determining the Bogoliubov energy $E_{\alpha \beta}$ and
the factors $v_{u,vw}^{\alpha \beta}$ within perturbation theory. It is required to perform the analysis up to second-order perturbation theory for 
$E_{\alpha \beta}$ and first order for $v_{u,vw}^{\alpha \beta}$ (see Supplemental Material). Then, the beyond-mean-field correction takes the form
\begin{equation}
  \frac{ E_{\rs{1D}}^{h}} {\lambda E^{h}_{0}} \overset{\kappa\ll1}{=}  \kappa^2  \frac{  C^{h}_{\rs{1D}}}{\sqrt{2}} - \frac{  4 \sqrt{2}  }{3\pi} \kappa^{3/2}  + \frac{4 \sqrt{2} \ln \left(4/3 \right)}{\pi} \kappa^{5/2}+   O(\kappa^3) . 
 \label{eq:harmonic1D}
\end{equation}
The first term on the right side accounts for the correction to the mean-field term due to the confinement-induced resonance 
with $C_{\rs{1D}}^{h} \approx 1.4603 $ \cite{Olshanii1998}, while the second  term describes the beyond-mean-field contribution 
of a purely  one-dimensional system.  Finally, the term with $\kappa^{5/2}$ provides the leading correction in the crossover. It is highly remarkable that for
harmonic confinement a term $\kappa^{5/2}$ appears, which was absent in the previous analysis with periodic boundary conditions. 
These predictions are fully confirmed with the numerical approach [see Fig.~\ref{fig2}(c)]. However, for a correct description of the zero crossing
it is also important to include the next term $B_{\rs{ 1D}}^{h} \kappa^3$ in the expansion. The prefactor $B_{\rs{1D}}^{h}$ is determined by a fitting procedure
to the numerical evaluation, which predicts  $B_{\rs{ 1D}}^{h} \approx 0.1$.

An analogous calculation can also be performed for the 3D-2D crossover within a harmonic confinement  $\kappa = n_{\rs{2D}} a_s l_{\perp}$.  Again, we expect
the prediction by local-density approximation for $\kappa\gg 1$, while for $\kappa\ll 1$ the ground-state energy reduces to the two-dimensional result 
$E_{\rs{2D}}^{h}= \hbar \omega_{\perp}  \kappa^2 \ln\left[ \kappa C_{\rs{2D}}^{h} \right]  L^2/l_{\perp}^2$
with $C_{\rs{2D}}^{h} \approx 28.69$ \cite{Petrov2001,Pricoupenko2008} including the renormalized scattering length. 
The next term due to the crossover  can again be derived within perturbation theory and provides a correction $\propto \kappa^3 \ln \kappa$.

{\it Conclusion.} 
We present a detailed study of the beyond-mean-field corrections for a weakly interacting Bose gas in the dimensional crossover.  While for a transverse 
confinement with periodic boundary conditions the analysis can be performed analytically, for a realistic setup with harmonic confinement a numerical 
analysis is required, and we find excellent agreement with the predictions from local-density approximation for $\kappa \gtrsim 1$. Furthermore,  we find that the 
correction to the local-density approximation lowers the ground-state energy.  This phenomenon might explain the recently observed systematic shift in the scattering 
length determined by the stability of self-bound droplets \cite{Schmitt2016}:   the finite extent of the droplets in transverse direction naturally introduces a confinement 
of the underlying gas and hence a correction to the local-density approximation.  In addition, our results show that the full crossover is excellently described by the combination 
of the leading contribution for $\kappa\ll1$  and $\kappa\gg1$, which in general is sufficient to describe the qualitative behavior  throughout the crossover. Our results are
immediately applicable to Bose-Bose mixtures of equal masses, as the treatment of the beyond-mean-field effects in the vicinity of the collapse reduces to that of a scalar 
gas with an effective interaction. Moreover, the results give a starting point for developing the beyond-mean-field theory in inhomogeneous systems with long-range interactions 
such as dipolar particles or Rydberg-dressed atoms. 

{\it Note added.}  During the final steps of preparation, we became aware of very related results by Zin {\it et al.} \cite{Zin2018}.

 {\bf Acknowledgments}
The research leading to these results received funding from the European Research Council (FR7/2007-2013 Grant Agreement No. 341197), and is supported by the Deutsche Forschungsgemeinschaft (DFG) within the research unit FOR 2247.

\bibliographystyle{apsrev4-1}
\input{paper-LHY-crossover.bbl}



\section*{Supplementary material}

# Supplemental Material: Dimensional crossover for the beyond-mean-field correction in Bose gases

Tobias Ilg,[1] Jan Kumlin,[1] Luis Santos,[2] Dmitry S. Petrov,[3] and Hans Peter Büchler[1]

[1]*Institute for Theoretical Physics III and Center for Integrated Quantum Science and Technology, University of Stuttgart, DE-70550 Stuttgart, Germany*
[2]*Institut für Theoretische Physik, Leibniz Universität Hannover, Appelstr. 2, DE-30167 Hannover, Germany*
[3]*LPTMS, CNRS, Univ. Paris Sud, Université Paris-Saclay, 91405 Orsay, France*
(Dated: November 28, 2018)

## BEYOND-MEAN-FIELD CORRECTION

In this paper, we use the field theoretic approach of Hugenholtz and Pines [1] to conveniently include the beyond-mean-field corrections for the weakly interacting Bose gas. In their paper, Hugenholtz and Pines are able to connect the ground state energy to the Green's function of the excited particles. The Green's function can then be evaluated using the standard procedure in quantum field theory and they obtain in beyond-mean-field order

$$E - \frac{1}{2}\mu N = \frac{1}{4}\sum_{\mathbf{k}\neq 0} \frac{2\tilde{\varepsilon}^2 + 3\tilde{\varepsilon}ng}{\sqrt{\tilde{\varepsilon}^2 + 2\tilde{\varepsilon}ng}} - 2\tilde{\varepsilon} - ng. \quad (1)$$

Here, $\tilde{\varepsilon} = \hbar^2 \mathbf{k}^2/2m$ is the single particle excitation spectrum, $g = 4\pi\hbar^2 a_s/m$ is the coupling constant, where $a_s$ denotes the scattering length, and $n = N/V$ is the density of the gas. The chemical potential satisfies the thermodynamic relation $\mu = dE/dN$. Hence, Eq. (1) is a linear first order differential equation for the ground state energy $E$. We now rewrite the differential equation in terms of the crossover parameter $\kappa$ and arrive at

$$E - \frac{\kappa}{2}\frac{dE}{d\kappa} = \frac{\lambda}{2}E_0 \sum\!\!\!\!\!\!\!\int \frac{2\varepsilon^2 + 3\varepsilon\kappa}{\sqrt{\varepsilon^2 + 2\varepsilon\kappa}} - 2\varepsilon - \kappa,$$

where $\varepsilon = \tilde{\varepsilon}/(4\pi\hbar^2/ml_\perp^2) = \pi(u^2 + v^2 + w^2)/2$ is the single particle excitation spectrum in dimensionless units and we have introduced the energy scale $E_0 = \frac{2\pi\hbar^2}{m}\frac{V}{l_\perp^4 a_s}$. We want to emphasize that this differential equation determines all terms except the mean-field terms of order $\kappa^2$. Hence, the divergencies from Bogoliubov theory appearing by a renormalization of the mean-field term are prevented. The general solution of this differential equation takes the from

$$\frac{E}{E_0} = \kappa^2\left[1 + \lambda\left(A(\kappa^*) - \int_{\kappa^*}^\kappa d\kappa'\, h(\kappa')\right)\right] \qquad \text{with} \qquad h(\kappa) = \frac{1}{\kappa^3}\sum\!\!\!\!\!\!\!\int\left[\frac{2\varepsilon^2 + 3\kappa\varepsilon}{\sqrt{\varepsilon^2 + 2\kappa\varepsilon}} - 2\varepsilon - \kappa\right].$$

The constant $A(\kappa^*)$ determines the initial condition of the differential equation and has to be chosen in order to reproduce the correct mean-field term proportional to $\kappa^2$, while $\kappa^*$ denotes an arbitrary value. The proper determination of $A(\kappa^*)$ is given by the condition to reproduce the correct ground state energy in Eq. (1) of the main text in the three-dimensional regime with $\kappa \gg 1$, which leads to

$$A(\kappa^*) = \int_{\kappa^*}^\infty [h(\kappa') - h_{3\mathrm{D}}(\kappa')] - \int_0^{\kappa^*} d\kappa' h_{3\mathrm{D}}(\kappa'). \quad (2)$$

This relation holds for all values of $\kappa^*$ so we choose $\kappa^* = \kappa$ and we arrive at $E/E_0 = \kappa^2(1 + \lambda A(\kappa))$. For large values of $\kappa$ the first term of Eq. (2) vanishes and the second exactly provides the correct three-dimensional LHY correction.

## EFFECTIVE SCATTERING LENGTH

We calculate the effective scattering lengths by making use of a two-channel model. In the open channel, we describe the two particles by the wave function $\psi(\mathbf{x}, \mathbf{y})$, whereas the closed channel is described by a single molecular state $\phi(\mathbf{z})$. The two channels are described by

$$\left[E - H_0^a - H_0^b\right]\psi(\mathbf{x}, \mathbf{y}) = \overline{g}\int d\mathbf{z}\,\alpha_\Lambda(\mathbf{r})\phi(\mathbf{z})\delta(\mathbf{z} - \mathbf{R}) \quad \text{and} \quad \left[E - \nu_0 - H_0^M\right]\phi(\mathbf{z}) = \overline{g}\int d\mathbf{x}d\mathbf{y}\,\alpha_\Lambda(\mathbf{r})\psi(\mathbf{x}, \mathbf{y})\delta(\mathbf{z} - \mathbf{R}), \quad (3)$$

where $\boldsymbol{r}$ and $\boldsymbol{R}$ are the relative and center-of-mass coordinate of the two particles, while $\alpha_\Lambda(\boldsymbol{r})$ describes the coupling between the two channels. Here, we introduce a cut-off $\Lambda$, such that $\alpha_\Lambda(\boldsymbol{r}) \to \delta(\boldsymbol{r})$ for $\Lambda \to 0$, e.g., $\alpha_\Lambda(\boldsymbol{r}) = e^{-\boldsymbol{r}^2/2\Lambda^2}/(2\pi\Lambda^2)^{3/2}$. Hence, we obtain a pseudopotential in the open channel for $\Lambda \to 0$ but a regular potential for $\Lambda \neq 0$. This regularization procedure of the pseudopotential is a suitable method to avoid divergencies throughout the calculation of the scattering amplitudes. The single particle Hamiltonians are given by $H_0^\sigma = -\frac{\hbar^2}{2m_\sigma}\Delta$ with $m_M = m_a + m_b$. In the following, we restrict ourselves to scattering processes with particles of equal mass $m_a = m_b = m$. To connect the three-dimensional scattering length $a_s$ to the effective lower-dimensional scattering lengths, we first need to solve the coupled Eq. (3) in the three-dimensional case. We transform both equations to the center-of-mass frame. In the center-of-mass frame, the molecular wave function is simply a constant $\phi_c$ and we obtain

$$\left[\frac{\hbar^2 \boldsymbol{k}^2}{m} + \frac{\hbar^2}{m}\Delta\right]\psi(\boldsymbol{r}) = \overline{g}\phi_c \alpha_\Lambda(\boldsymbol{r}) \qquad \text{and} \qquad \left[\frac{\hbar^2 \boldsymbol{k}^2}{m} - \nu_0\right]\phi_c = \overline{g}\int d\boldsymbol{r}\,\psi(\boldsymbol{r})\alpha_\Lambda(\boldsymbol{r}). \tag{4}$$

For the remaining Eq. (4) we choose the ansatz

$$\psi_{\boldsymbol{k}}(\boldsymbol{r}) = e^{\mathrm{i}\boldsymbol{k}\boldsymbol{r}} + \beta \int d\boldsymbol{r}'\, \alpha_\Lambda(\boldsymbol{r}') G_{\boldsymbol{k}}(\boldsymbol{r}-\boldsymbol{r}'), \tag{5}$$

where we define the Green's function by $\frac{\hbar^2}{m}\left[\boldsymbol{k}^2 + \Delta\right]G_{\boldsymbol{k}}(\boldsymbol{r}) = \delta(\boldsymbol{r})$. The far-field behavior of our ansatz yields a connection between the factor $\beta$ and the low-energy scattering amplitude,

$$f(\boldsymbol{k}) = -\frac{m}{4\pi\hbar^2}\beta = -\frac{m}{4\pi\hbar^2}\frac{\overline{g}^2}{\frac{\hbar^2 \boldsymbol{k}^2}{m} - \nu_0 - \overline{g}^2 \overline{G}(0) + \frac{m\overline{g}^2}{4\pi\hbar^2}\mathrm{i}k} \qquad \text{where} \qquad \overline{G}(0) = \int d\boldsymbol{r}\, d\boldsymbol{r}'\, \alpha_\Lambda(\boldsymbol{r})\alpha_\Lambda(\boldsymbol{r}') G_0(\boldsymbol{r}-\boldsymbol{r}').$$

Thus, the scattering amplitude simplifies to

$$f(\boldsymbol{k}) = -\frac{1}{-\frac{4\pi\hbar^2}{m}\frac{\nu}{\overline{g}^2} + \mathrm{i}k + \mathcal{O}(k^2)} = -\frac{1}{\frac{1}{a_s} + \mathrm{i}k},$$

where we have introduced the renormalized detuning $\nu = \nu_0 + g^2 \overline{G}(0)$. Hence, we have found the connection between the scattering length and the parameters of the two-channel model $a_s = -\frac{m}{4\pi\hbar^2}\frac{\overline{g}^2}{\nu}$.

**Quasi one-dimensional scattering**

With the discussed procedure, we can now treat confined systems analogously and connect the effective scattering length $a_{1\mathrm{D}}$ to the three-dimensional scattering length $a_s$. We will begin with a quasi one-dimensional geometry. We impose periodic boundary conditions and the allowed wave vectors are of the form $k_z \in \mathbb{R}$ in the elongated direction and $k_x = \frac{2\pi}{l_\perp}v$, $k_y = \frac{2\pi}{l_\perp}w$ with $v,w \in \mathbb{Z}$ in the transversal directions. As we are interested in the quasi one-dimensional case, the particles occupy the transverse ground state ($v = w = 0$) and the kinetic energy along the $z$-direction is much smaller than the level spacing, $k_z \ll 2\pi/l_\perp$. With the restrictions on the wave vectors in mind, the coupled Eq. (4) remain valid and so does our ansatz Eq. (5). Again, it is possible to find a connection between the quasi one-dimensional scattering amplitude and the factor $\beta_{1\mathrm{D}}$ using the behavior of the wave function at large interparticle separations,

$$f_{1\mathrm{D}}(k_z) = \frac{\mathrm{i}}{2}\frac{m}{\hbar^2 l_\perp^2}\beta_{1\mathrm{D}} = \frac{\mathrm{i}}{2}\frac{m}{\hbar^2 l_\perp^2}\frac{\overline{g}^2}{\frac{\hbar^2 \boldsymbol{k}^2}{m} - \nu_0 - \overline{g}^2 \overline{G}_{1\mathrm{D}}(k_z)},$$

where

$$\overline{G}_{1\mathrm{D}}(k_z) = \int d\boldsymbol{r}\, d\boldsymbol{r}'\, \alpha_\Lambda(\boldsymbol{r})\alpha_\Lambda(\boldsymbol{r}') G_{k_z}^{1\mathrm{D}}(\boldsymbol{r}-\boldsymbol{r}') = \frac{m}{\hbar^2}\frac{1}{l_\perp^2}\sum_{p_x,p_y}\int \frac{dp_z}{2\pi}\frac{\hat{\alpha}_\Lambda^2(\boldsymbol{p})}{k_z^2 - p^2 + \mathrm{i}\eta}$$

$$= -\frac{\mathrm{i}m}{2\hbar^2 l_\perp^2 k_z} + \frac{m}{\hbar^2 l_\perp^2}\sum_{(p_x,p_y)\neq(0,0)}\int \frac{dp_z}{2\pi}\frac{\hat{\alpha}_\Lambda^2(\boldsymbol{p})}{k_z^2 - p^2 + \mathrm{i}\eta} = -\frac{\mathrm{i}m}{2\hbar^2 l_\perp^2 k_z} + \overline{G}'_{1\mathrm{D}}(k_z).$$

Here, $\hat{\alpha}_\Lambda(\boldsymbol{p})$ denotes the Fourier transformed of $\hat{\alpha}_\Lambda(\boldsymbol{r})$ and we separated the mode $p_x = p_y = 0$ in the last step. Finally, we arrive at an expression for the quasi one-dimensional scattering amplitude

$$f_{1\mathrm{D}}(k_z) = -\frac{1}{1 + \mathrm{i}k_z \left(\frac{2\hbar^2 l_\perp^2}{m}\left(\frac{\nu}{\bar{g}^2} + \overline{G}'_{1\mathrm{D}}(0) - \overline{G}(0)\right)\right) + O(k_z^2)} = -\frac{1}{1 + \mathrm{i}k_z a_{1\mathrm{D}}},$$

where the effective scattering length $a_{1\mathrm{D}}$ is given by

$$a_{1\mathrm{D}} = -\frac{1}{2\pi}\frac{l_\perp^2}{a_s}\left(1 - C_{1\mathrm{D}}\frac{a_s}{l_\perp}\right) \qquad \text{with} \qquad C_{1\mathrm{D}} = \int dv dw \frac{1}{\sqrt{v^2 + w^2}} - {\sum_{v,w}}' \frac{1}{\sqrt{v^2 + w^2}} \approx 3.899.$$

This expression describes the confinement-induced resonance for a setup with periodic boundary conditions.

### Quasi two-dimensional scattering

Next, we calculate the effective two-dimensional scattering length $a_{2\mathrm{D}}$ in the same manner. The allowed wave vectors are of the form $k_y, k_z \in \mathbb{R}$ and $k_x = \frac{2\pi}{l_\perp} w$ with $w \in \mathbb{Z}$. The motion is restricted to the transverse ground state and an analysis of the far-field behavior of the given geometry gives an expression for the scattering amplitude

$$f_{2\mathrm{D}}(k_\rho) = -\frac{m}{4\hbar^2}\sqrt{\frac{2\mathrm{i}}{\pi k_\rho}}\frac{1}{l_\perp}\beta_{2\mathrm{D}} = -\frac{m}{4\hbar^2}\sqrt{\frac{2\mathrm{i}}{\pi k_\rho}}\frac{1}{l_\perp}\frac{\bar{g}^2}{\frac{\hbar^2 k_\rho^2}{m} - \nu_0 - \bar{g}^2 \overline{G}_{2\mathrm{D}}(k_\rho)},$$

where we have introduced $k_\rho^2 = k_y^2 + k_z^2$, and

$$\overline{G}_{2\mathrm{D}}(k_\rho) = \frac{m}{\hbar^2}\frac{1}{l_\perp}\sum_{p_x}\int \frac{dp_y dp_z}{(2\pi)^2}\frac{\hat{\alpha}_\Lambda^2(\boldsymbol{p})}{k_\rho^2 - p^2 + \mathrm{i}\eta}.$$

Replacing the bare detuning $\nu_0$ with the physical detuning $\nu$, we arrive at

$$f_{2\mathrm{D}}(k_\rho) = -\sqrt{\frac{2\pi\mathrm{i}}{k_\rho}}\frac{1}{\frac{l_\perp}{a_s} - \frac{4\pi\hbar^2 l_\perp}{m}\left(\overline{G}^{2\mathrm{D}}(k_\rho) - \overline{G}(0)\right)} = -\sqrt{\frac{2\pi\mathrm{i}}{k_\rho}}\frac{1}{\mathrm{i}\pi - \ln\left(k_\rho^2 l_\perp^2 e^{-l_\perp/a_s}\right)} = -\sqrt{\frac{2\pi\mathrm{i}}{k_\rho}}\frac{1}{\mathrm{i}\pi - \ln\left(k_\rho^2 a_{2\mathrm{D}}^2 e^{2\gamma}/4\right)},$$

where $\gamma \approx 0.577$ is the Euler-Mascheroni constant, and we have used the relation

$$\overline{G}^{2\mathrm{D}}(k_\rho) - \overline{G}(0) = \frac{m}{4\pi\hbar^2 l_\perp}\left(-\mathrm{i}\pi + \ln(k_\rho^2 l_\perp^2)\right).$$

Hence, the effective scattering length in the quasi two-dimensional geometry is given by $a_{2\mathrm{D}} = 2l_\perp e^{-\frac{l_\perp}{2a_s}}e^{-\gamma}$.

### Harmonic confinement

In this part, we provide a derivation of the effective two-dimensional scattering length $a_{2\mathrm{D}}$ in a setup with harmonic confinement. Again, we resort to the previously studied two-channel model. The introduction of a harmonic trapping potential in $x$-direction changes the single particle Hamiltonian to $H_0 = -\frac{\hbar^2}{2m}\Delta + \frac{m\omega^2}{2}x^2$ and gives rise to the harmonic oscillator length $l_\perp = \sqrt{\hbar/m\omega_\perp}$. The coupled Schrödinger equations in the center-of-mass frame reduce to

$$\left[\frac{\hbar^2 \boldsymbol{k}_\rho^2}{m} + \hbar\omega_\perp\left(n + \frac{1}{2}\right) + \frac{\hbar^2}{m}\Delta - \frac{m\omega_\perp^2}{4}x^2\right]\psi(\boldsymbol{r}) = \bar{g}\phi_c \alpha_\Lambda, \text{ and } \left[\frac{\hbar^2 \boldsymbol{k}_\rho^2}{m} + \hbar\omega_\perp\left(n + \frac{1}{2}\right) - \nu_0\right]\phi_c = \bar{g}\int d\boldsymbol{r}\, \alpha_\Lambda(\boldsymbol{r})\psi(\boldsymbol{r}),$$

where $\boldsymbol{k}_\rho = (k_y, k_z)^T$. We are interested in the quasi two-dimensional regime so the motion is restricted to the transverse ground state and our ansatz for the scattering wave function takes the form

$$\psi(\boldsymbol{r}) = e^{\mathrm{i}\boldsymbol{k}_\rho \boldsymbol{\rho}}\varphi_0(x) + \beta_{2\mathrm{D}}^h \int d\boldsymbol{r}'\, \alpha_\Lambda(\boldsymbol{r}') G_{k_\rho,0}(\boldsymbol{r}, \boldsymbol{r}')$$

with $\boldsymbol{\rho} = (y, z)^T$. In addition, $\varphi_n(x)$ denote the eigenfunctions of the harmonic oscillator. Then, the Green's function is given by

$$G_{k_\rho,0}(\boldsymbol{r},\boldsymbol{r}') = \frac{m}{\hbar^2} \int \frac{d\boldsymbol{p}^2}{(2\pi)^2} \sum_{n=0}^{\infty} \frac{e^{i\boldsymbol{p}(\boldsymbol{\rho}-\boldsymbol{\rho}')}\varphi_n(x)\varphi_n(x')}{\boldsymbol{k}_\rho^2 - \boldsymbol{p}^2 - n/l_\perp^2 + i\eta},$$

Again, we find the connection to the quasi two-dimensional scattering amplitude $f_{2D}^h(\boldsymbol{k}_\rho)$,

$$f_{2D}^h(\boldsymbol{k}_\rho) = -\frac{m}{4\hbar^2}\sqrt{\frac{2i}{\pi k_\rho}}\left(\frac{1}{2\pi l_\perp^2}\right)^{1/4}\beta_{2D}^h = -\frac{m}{4\hbar^2}\sqrt{\frac{2i}{\pi k_\rho}}\left(\frac{1}{2\pi l_\perp^2}\right)^{1/4}\frac{\bar{g}^2\left(\frac{1}{2\pi l_\perp^2}\right)^{1/4}}{\frac{\hbar^2 \boldsymbol{k}_\rho^2}{m} - \nu_0 - \bar{g}^2 \overline{G}_{2D}^h(\boldsymbol{k}_\rho)},$$

where

$$\overline{G}_{2D}^h(\boldsymbol{k}_\rho) = \int d\boldsymbol{r}\, d\boldsymbol{r}'\, \alpha_\Lambda(\boldsymbol{r})\alpha_\Lambda(\boldsymbol{r}') G_{k_\rho,n=0}(\boldsymbol{r},\boldsymbol{r}').$$

Thus, the quasi two-dimensional scattering amplitude in terms of the physical detuning $\nu$ simplifies to

$$f_{2D}^h(\boldsymbol{k}_\rho) = -\frac{m}{4\hbar^2}\sqrt{\frac{2i}{\pi k_\rho}}\frac{1}{\sqrt{2\pi l_\perp^2}}\frac{\bar{g}^2}{-\nu - \bar{g}^2\left(\overline{G}_{2D}^h(\boldsymbol{k}_\rho) - \overline{G}(0)\right)}.$$

For the evaluation of the difference $\overline{G}_{2D}^h(\boldsymbol{k}_\rho) - \overline{G}(0)$ it is suitable to choose $\alpha_\Lambda(\boldsymbol{r}) = e^{-\rho/2\Lambda^2}\delta(x)/2\pi\Lambda^2$ without loss of generality. Then, we can analytically evaluate the summation over the harmonic oscillator modes in $\overline{G}_{2D}^h(\boldsymbol{k}_\rho)$ and are left with the integration in momentum space,

$$\overline{G}_{2D}^h(\boldsymbol{k}_\rho) - \overline{G}(0) = \frac{m}{4\pi\hbar^2}\frac{1}{\sqrt{2\pi l_\perp^2}}\left(\overline{C}_{2D}^h - i\pi + \ln\left(\boldsymbol{k}_\rho^2 l_\perp^2/2\right)\right),$$

where

$$\overline{C}_{2D}^h = \int_0^\infty du\, 2u\left(\frac{\sqrt{\pi}}{u} - \frac{\sqrt{\pi}\Gamma(u^2)}{\Gamma(u^2 + \frac{1}{2})} + \frac{1}{u^2(1+u)}\right) \approx 1.938\,.$$

Hence, the scattering amplitude $f_{2D}^h(\boldsymbol{k}_\rho)$ is

$$f_{2D}^h(\boldsymbol{k}_\rho) = -\sqrt{\frac{2\pi i}{k_\rho}}\frac{1}{i\pi - \ln\left(\boldsymbol{k}_\rho^2 l_\perp^2 e^{\overline{C}_{2D}^h - \sqrt{2\pi}l_\perp/a_s}\right)} = -\sqrt{\frac{2\pi i}{k_\rho}}\frac{1}{i\pi - \ln\left(\boldsymbol{k}_\rho^2 a_{2D}^{h\,2} e^{2\gamma}/4\right)},$$

and we found an expression for the scattering length, $a_{2D}^h = \sqrt{2}l_\perp e^{-\gamma - \overline{C}_{2D}^h/2} e^{-\sqrt{2\pi}l_\perp/2a_s}$. A closer analysis of $\overline{C}_{2D}^h$ reveals the connection to the constant $B$ found in [2, 3], $B = 2\pi e^{-\overline{C}_{2D}^h} \approx 0.905$, and we obtain

$$a_{2D}^h = 2l_\perp \sqrt{\frac{\pi}{B}} e^{-\sqrt{2\pi}l_\perp/2a_s - \gamma}.$$

The constant $C_{2D}^h$ of the main text is then given by $C_{2D}^h = (2\pi)^{3/2}\sqrt{e}/B$. One can obtain the quasi one-dimensional scattering length in a harmonic confinement in the same manner and for completeness we will only give the expression first found by [4] here,

$$a_{1D}^h = -\frac{l_\perp^2}{a_s}\left(1 - \frac{C_{1D}^h}{\sqrt{2}}\frac{a_s}{l_\perp}\right) \qquad \text{with} \qquad C_{1D}^h \approx 1.4603\,.$$

### 3D-1D CROSSOVER

In the following, we will briefly sketch the derivation of the function describing the entire crossover regime in the 3D-1D crossover. We start with

$$\frac{E - E_{3D}}{E_0} = \kappa^2 \lambda \int_\kappa^\infty d\kappa'\, [h(\kappa') - h_{3D}(\kappa')]$$

and perform the integration over $\kappa'$ first. The upper boundary of the integration vanishes and we obtain

$$\frac{E_{1D} - E_{3D}}{E_0} = \lambda \int du \left[ \sum_{n,m} f(\varepsilon) - \int dv dw\, f(\varepsilon) \right], \qquad \text{where} \qquad f(\varepsilon) = \sqrt{\varepsilon^2 + 2\varepsilon\kappa} - \varepsilon - \kappa.$$

We proceed by rewriting the integration over $u$ by deforming the integration path into the complex plane. Then, the integral takes the from

$$\int_{-\infty}^{\infty} du \sqrt{\varepsilon^2 + 2\varepsilon\kappa} - \varepsilon - \kappa = -\frac{4\kappa^{3/2}}{\sqrt{\pi}} \int_0^1 dt\, \frac{\sqrt{t(1-t)}}{\sqrt{t + \frac{\pi}{4\kappa}(v^2 + w^2)}}.$$

This allows us now to nicely separate the different contributions to the ground state energy,

$$\int dv\,dw \frac{1}{\sqrt{t + \frac{\pi}{4\kappa}(v^2 + w^2)}} - \sum_{v,w} \frac{1}{\sqrt{t + \frac{\pi}{4\kappa}(v^2 + w^2)}}$$

$$\left. \begin{aligned} &= -\frac{1}{\sqrt{t}} + {\sum_{v,w}}' \frac{1}{\sqrt{\frac{\pi}{4\kappa}(v^2 + w^2)}} - {\sum_{v,w}}' \frac{1}{\sqrt{t + \frac{\pi}{4\kappa}(v^2 + w^2)}} \\ &\quad + \int dv\,dw \frac{1}{\sqrt{\frac{\pi}{4\kappa}(v^2 + w^2)}} - {\sum_{v,w}}' \frac{1}{\sqrt{\frac{\pi}{4\kappa}(v^2 + w^2)}} \end{aligned} \right\} = \sqrt{\frac{4\kappa}{\pi}} C_{1D}$$

$$\left. + \int dv\,dw \int \frac{1}{\sqrt{t + \frac{\pi}{4\kappa}(v^2 + w^2)}} - \int dv\,dw \frac{1}{\sqrt{\frac{\pi}{4\kappa}(v^2 + w^2)}} \right\} = -8\sqrt{t}\kappa.$$

Hence, we arrive at

$$\frac{E - E_{3D}}{\lambda E_0} = -\frac{8}{3\sqrt{\pi}} \kappa^{3/2} + C_{1D}\kappa^2 - \frac{128}{15\sqrt{\pi}} \kappa^{5/2} + \frac{8\kappa^2}{\pi} \int_0^1 dt\, \sqrt{t(1-t)} \left[ {\sum_{v,w}}' \frac{1}{\sqrt{v^2 + w^2}} - {\sum_{v,w}}' \frac{1}{\sqrt{v^2 + w^2 + \frac{4\kappa t}{\pi}}} \right].$$

As a last step, we want to get rid of the double summation. Therefore, we use $1/\sqrt{A} = \int_0^\infty d\tau\, e^{-\tau A}/\sqrt{\tau}$ to write

$$\begin{aligned} {\sum_{v,w}}' \left[ \frac{1}{\sqrt{v^2 + w^2}} - \frac{1}{\sqrt{v^2 + w^2 + \frac{4\kappa t}{\pi}}} \right] &= \int_0^\infty \frac{d\tau}{\sqrt{\pi\tau}} {\sum_{v,w}}' e^{-\tau(v^2 + w^2)} (1 - e^{-\tau 4\kappa t/\pi}) \\ &= \int_0^\infty \frac{d\tau}{\sqrt{\pi\tau}} \left( \vartheta_3(0, e^{-\tau})^2 - 1 \right) \left( 1 - e^{-\tau 4\kappa t/\pi} \right), \end{aligned}$$

where we have made use of the definition of the Jacobi theta function $\vartheta_3(z, q) = \sum_n q^{n^2} \cos(2nz)$. Finally, we perform the integration over $t$,

$$\int_0^1 dt \sqrt{t(1-t)} \left( 1 - e^{-4\tau\kappa t/\pi} \right) = \frac{\pi}{8} \left( 1 - \frac{e^{-\frac{2\tau\kappa}{\pi}} I_1\left(\frac{2\tau\kappa}{\pi}\right)}{\tau\kappa/\pi} \right),$$

and arrive at the final result

$$\frac{E_{1D}}{E_0} = \kappa^2(1 + \lambda C_{1D}) - \lambda \frac{8}{3\sqrt{\pi}} \kappa^{3/2} + \lambda \kappa^2 \int_0^\infty \frac{d\tau}{\sqrt{\pi\tau}} \left[ \vartheta_3\left(0, e^{-\tau}\right)^2 - 1 \right] \left[ 1 - \frac{e^{-\frac{2\tau\kappa}{\pi}} I_1\left(\frac{2\tau\kappa}{\pi}\right)}{\tau\kappa/\pi} \right].$$

### 3D regime with $\kappa \gg 1$

Using the relation $\vartheta_3\left(0, e^{-\pi x}\right) = \vartheta_3\left(0, e^{-\pi/x}\right)/\sqrt{x}$, which is straightforward to prove using the Poisson summation formula, we find a suitable expression to perform the analytic expansion for $\kappa \gg 1$,

$$\left.\begin{aligned}&\kappa^2\int_0^\infty\frac{d\tau}{\sqrt{\pi\tau}}\left[\vartheta_3\left(0,e^{-\tau}\right)^2-1\right]\left[1-\frac{e^{-\frac{2\tau\kappa}{\pi}}I_1\left(\frac{2\tau\kappa}{\pi}\right)}{\tau\kappa/\pi}\right]=\kappa^2\int_0^\infty\frac{d\tau}{\sqrt{\pi\tau}}\left[\vartheta_3\left(0,e^{-\tau}\right)^2-\frac{\pi}{\tau}\right]\left[1-\frac{e^{-\frac{2\pi\kappa}{\tau}}I_1\left(\frac{2\pi\kappa}{\tau}\right)}{\pi\kappa/\tau}\right]\\ &=\kappa^2\int_0^\infty\frac{d\tau}{\sqrt{\pi\tau}}\left[1-\frac{e^{-\frac{2\pi\kappa}{\tau}}I_1\left(\frac{2\pi\kappa}{\tau}\right)}{\pi\kappa/\tau}\right]\end{aligned}\right\}=\frac{128}{15\sqrt{\pi}}\kappa^{5/2} \quad (6)$$

$$\left.+\kappa^2\int_0^\infty\frac{d\tau}{\sqrt{\pi\tau}}\left[\vartheta_3\left(0,e^{-\tau}\right)^2-\frac{\pi}{\tau}-1\right]\right\}=-\kappa^2 C_{1D} \quad (7)$$

$$\left.+\kappa^2\int_0^\infty\frac{d\tau}{\sqrt{\pi\tau}}\frac{\pi}{\tau}\left[\frac{e^{-\frac{2\pi\kappa}{\tau}}I_1\left(\frac{2\pi\kappa}{\tau}\right)}{\pi\kappa/\tau}\right]\right\}=\frac{8}{3\sqrt{\pi}}\kappa^{3/2} \quad (8)$$

$$-\int_0^\infty\frac{d\tau}{\sqrt{\pi\tau}}\left[\vartheta_3\left(0,e^{-\tau}\right)^2-1\right]\left[\frac{e^{-\frac{2\pi\kappa}{\tau}}I_1\left(\frac{2\pi\kappa}{\tau}\right)}{\pi\kappa/\tau}\right]\approx-\kappa^2\int_0^\infty\frac{d\tau}{\sqrt{\pi\tau}}\left[\vartheta_3\left(0,e^{-\tau}\right)^2-1\right]\sqrt{\frac{2}{\pi}}\left(\frac{\tau}{2\kappa\pi}\right)^{3/2}=-\sqrt{\kappa}\frac{A_{1D}}{2\pi^{5/2}}$$

Inserting into Eq. (9) of the main text, we see that Eq. (6) provides the correct 3D results, while Eq. (7) cancels the correction to the mean-field shift, Eq. (8) cancels the 1D beyond-mean-field correction, while the last term can now be expanded in the small parameter $\tau/\kappa$, which provides the leading correction for large $\kappa\gg 1$ with $A_{1D}=\sum'_{v,w}\left(v^2+w^2\right)^{-2}\approx 6.0268$.

### HARMONIC CONFINEMENT

**Bogoliubov theory in the 1D geometry with harmonic confinement**

In the following, we study a system, where the mean-field energy is canceled by a second type of interaction, and the ground state remains always in the lowest harmonic mode of the transverse confinement within mean-field theory. Then, we can derive the Bogoliubov excitation spectrum in analogy to the situation with periodic boundary conditions and perform a perturbation expansion for small values of $\kappa=n_{1D}a_s$. We express the bosonic field operator in eigenstates of the non-interacting theory,

$$\psi(\boldsymbol{r})=\frac{1}{\sqrt{L}}\sum_{v,w}\int\frac{dk}{2\pi}e^{\mathrm{i}kx}\varphi_v(y)\varphi_w(z)b_{k,vw} \quad\text{with}\quad \varphi_w(z)=\frac{1}{\sqrt{2^w w!\,l_\perp\sqrt{\pi}}}H_w(z/l_\perp)e^{-z^2/2l_\perp^2}$$

the eigenfunctions of the harmonic oscillator and the oscillator length $l_\perp=\sqrt{\hbar/m\omega_\perp}$. The quantum many-body Hamiltonian takes the form

$$H=\sum_k\sum_{v,w}\tilde{\epsilon}_{k,vw}b^\dagger_{k,vw}b_{k,vw}+\frac{1}{2L}\sum_{\substack{k,k',\\q}}\sum_{\substack{i,i',w,w',\\j,j',v,v'}}V^{ij,i'j'}_{vw,v'w'}(q)b^\dagger_{k,vw}b^\dagger_{k',v'w'}b_{k'-q,i'j'}b_{k+q,vw}$$

with the excitation spectrum $\tilde{\epsilon}_{k,vw}=\frac{\hbar^2 k^2}{2m}+\hbar\omega_\perp(v+w)$ and the interaction potential

$$V^{ij,i'j'}_{vw,v'w'}(q)=\int dx\,dy\,dy'\,dz\,dz'\,e^{iqx}\varphi_v(y)\varphi_{v'}(y')\varphi_w(z)\varphi_{w'}(z')\varphi_{i'}(y')\varphi_i(y)\varphi_j(z)\varphi_{j'}(z')V(x,y,y',z,z').$$

In order to achieve the cancellation of the mean-field energy, we add an attractive interaction with a very long range. The combined interaction potential is suitably chosen as

$$V(x,y,y',z,z')=g\delta(y-y')\delta(z-z')\left(\delta(x)-\frac{1}{\sqrt{\pi r_0^2}}e^{-x^2/r_0^2}\right).$$

The first term is the contact interaction with $g=4\pi\hbar^2 a_s/m$, whereas the attractive second part will not contribute to the beyond-mean-field corrections due to its long-range character. The mean-field part, however, is strongly influenced by the additional interaction, as $\mu=V^{00,00}_{00,00}(0)=0$. Following the standard approach by Bogoliubov,

we replace the lowest state $b_{0,00}$ by a macroscopic occupation $\sqrt{N_0}$. Then, we express the condensate fraction by $N_0 = N - \sum'_{k,vw} b^\dagger_{k,vw} b_{k,vw}$ and obtain in leading order in $N$

$$H = \sum_k \sum_{vw} \tilde{\epsilon}_{k,vw} b^\dagger_{k,vw} b_{k,vw} + \frac{n_{1D}}{2} {\sum_{k,v,w,v',w'}}' \left[ 2V_{vw,v'w'} b^\dagger_{k,vw} b_{k,v'w'} + V_{vw,v'w'} b^\dagger_{k,vw} b^\dagger_{-k,v'w'} + V_{vw,v'w'} b_{k,vw} b_{-k,v'w'} \right],$$

where we have introduced $V_{vw,v'w'} = g \int dy\, dz\, \varphi_v(y)\varphi_{v'}(y)\varphi_0(y)^2 \varphi_w(z)\varphi_{w'}(z)\varphi_0(z)^2$ and the primed sum indicates the absence of the condensate mode. The determination of the Bogoliubov excitation spectrum is most conveniently achieved using the following approach: We start with the Heisenberg equation

$$i\hbar \dot{b}_{k,vw} = [b_{k,vw}, H] = \tilde{\epsilon}_{k,vw} b_{k,vw} + n_{1D} \sum_{v',w'} V_{vw,v'w'} (b_{k,v'w'} + b^\dagger_{-k,v'w'}).$$

Then, the second time derivation simplifies to

$$(i\hbar)^2 \ddot{b}_{k,vw} = \tilde{\epsilon}^2_{k,vw} b_{k,vw} + \tilde{\epsilon}_{k,vw} n_{1D} \sum_{v'w'} V_{vw,v'w'} \left( b_{k,v'w'} + b^\dagger_{-k,v'w'} \right) + n_{1D} \sum_{v'w'} V_{vw,v'w'} \left( \tilde{\epsilon}_{k,v'w'} b_{k,v'w'} - \tilde{\epsilon}_{k,v'w'} b^\dagger_{-k,v'w'} \right). \tag{9}$$

Adding and subtracting the adjoint of Eq. (9) and making use of the Bogoliubov transformation $b_{k,vw} = \sum_{\alpha,\beta} u^{\alpha\beta}_{k,vw} a_{k,\alpha\beta} + v^{\alpha\beta}_{k,vw} a^\dagger_{-k,\alpha\beta}$, we obtain two equations for the excitation spectrum $\tilde{E}_{k,\alpha\beta}$,

$$\tilde{E}^2_{k,\alpha\beta} f^{+,\alpha\beta}_{k,vw} = \tilde{\epsilon}^2_{k,vw} f^{+,\alpha\beta}_{k,vw} + 2\tilde{\epsilon}_{k,vw} n_{1D} \sum_{v'w'} V_{vw,v'w'} f^{+,\alpha\beta}_{k,v'w'} \quad \text{and}$$

$$\tilde{E}^2_{k,\alpha\beta} f^{-,\alpha\beta}_{k,vw} = \tilde{\epsilon}^2_{k,vw} f^{-,\alpha\beta}_{k,vw} + 2 n_{1D} \sum_{v'w'} \tilde{\epsilon}_{k,v'w'} V_{v'w',vw} f^{-,\alpha\beta}_{k,v'w'}, \tag{10}$$

with $f^{\pm,\alpha\beta}_{k,vw} = u^{\alpha\beta}_{k,vw} \pm v^{\alpha\beta}_{k,vw}$. Both equations are connected by the relation $f^{+,\alpha\beta}_{k,vw} = \frac{\tilde{\epsilon}_{k,vw}}{\tilde{E}_{k,\alpha\beta}} f^{-,\alpha\beta}_{k,vw}$. As the Bogoliubov transformation has to be canonical, the Bogoliubov functions satisfy $\delta_{\alpha,\gamma}\delta_{\beta,\delta} = \sum_{vw} f^{+,\alpha\beta}_{k,vw} f^{-,\gamma\delta}_{k,vw}$, which determines the normalization of the Bogoliubov functions $f^{\pm,\alpha\beta}_{k,vw}$. The solution of Eq. (10) provides the excitation spectrum $\tilde{E}_{k,\alpha\beta}$, the Bogoliubov function $f^{\pm,\alpha\beta}_{k,vw}$ and hence the amplitude $v^{\alpha\beta}_{k,vw}$. This allows us to determine the beyond-mean-field correction by the approach of Hugenholtz and Pines,

$$E^h_{1D} - \frac{1}{2}\mu N = \frac{L}{2} \int \frac{dk}{2\pi} \sum_{\substack{v,\alpha \\ w,\beta}} \left[ \tilde{\epsilon}_{k,vw} - \tilde{E}_{k,\alpha\beta} \right] |v^{\alpha\beta}_{k,vw}|^2. \tag{11}$$

In general, Eq.(10) has to be solved numerically. In the quasi-one-dimensional regime $\kappa = n_{1D} a_s \ll 1$, however, we can perform a perturbation expansion for $E_{k,\alpha\beta}$ and $f^{\pm,\alpha\beta}_{k,vw}$ in $\kappa$ to obtain the leading contributions. A consistent expansion in $\kappa$ turns out to be challenging and we will describe the procedure in detail. We introduce the dimensionless single particle excitation spectrum $\epsilon_{u,vw} = \tilde{\epsilon}_{k,vw}/(4\pi\hbar^2/ml_\perp^2) = \pi(u^2 + (v+w)/(2\pi^2))/2$ and the dimensionless Bogoliubov spectrum $E_{u,\alpha\beta} = \tilde{E}_{k,\alpha\beta}/(4\pi\hbar^2/ml_\perp^2)$, where $u = kl_\perp/(2\pi)$. In what follows, we will discuss the right-hand side of Eq. (11) and separate the different contributions,

$$\frac{L}{2} \int \frac{dk}{2\pi} \sum_{\substack{v,\alpha \\ w,\beta}} \left[ \tilde{\epsilon}_{k,vw} - \tilde{E}_{k,\alpha\beta} \right] |v^{\alpha\beta}_{k,vw}|^2 = 2\pi E^h_0 \lambda \Bigg( \int du\, [\epsilon_{u,00} - E_{u,00}] |v^{00}_{u,00}|^2 + \int du\, {\sum_{v,w}}' [\epsilon_{u,vw} - \epsilon_{u,00}] |v^{00}_{u,vw}|^2 \tag{12}$$

$$+ \int du\, {\sum_{\alpha,\beta}}' [\epsilon_{u,00} - E_{u,\alpha\beta}] |v^{\alpha\beta}_{u,00}|^2 + \int du\, {\sum_{v,w}}' {\sum_{\alpha,\beta}}' [\epsilon_{u,vw} - E_{u,\alpha\beta}] |v^{\alpha\beta}_{u,vw}|^2 \Bigg),$$

where we have introduced the energy scale $E^h_0 = \hbar\omega_\perp L/a_s$ and $\lambda = a_s/l_\perp$. For the first term we determine the Bogoliubov spectrum within second order perturbation theory,

$$E^2_{u,00} = \epsilon^2_{u,00} + 2\epsilon_{u,00} \kappa \eta_{00} \eta_{00} + 4\kappa^2 {\sum_{v,w}}' \frac{\epsilon_{u,00} \epsilon_{u,vw} \eta^2_{v0} \eta^2_{w0}}{\epsilon^2_{u,00} - \epsilon^2_{u,vw}},$$

where we have introduced the overlap of the harmonic oscillator wave functions

$$\eta_{vw} = l_\perp \int dz\, \varphi_v(z) \varphi_w(z) \varphi_0(z)^2 = \begin{cases} \frac{1}{\sqrt{2\pi}} \frac{(-1)^{(v-w)/2}}{2^{v+w}} \frac{(v+w)!}{(\frac{v+w}{2})!} \frac{1}{\sqrt{v!w!}} & v+w = \text{even} \\ 0 & v+w = \text{odd}. \end{cases}$$

The momentum dependence of the third term can be neglected, and we can evaluate the remaining sum over the harmonic oscillator modes,

$$\sum_{v,w}{}' \frac{\eta_{v0}^2 \eta_{w0}^2}{v+w} = \frac{\ln(4/3)}{8\pi^2}. \tag{13}$$

Thus, the Bogoliubov spectrum reads

$$E_{u,00}^2 = \epsilon_{u,00}^2 + 2\epsilon_{u,00}\left(\frac{\kappa}{2\pi} - \frac{\ln(4/3)}{\pi}\kappa^2\right).$$

For the Bogoliubov function $f_{u,00}^{+,00}$ on the other hand, it is sufficient only to include the lowest order $(f_{u,00}^{+,00})^2 = \frac{\epsilon_{u,00}}{E_{u,00}}$. Any higher order will only provide contribution of order $\kappa^{7/2}$ or higher. We obtain the amplitude

$$|v_{u,00}^{00}|^2 = \frac{1}{4}\left(1 - \frac{E_{u,00}}{\epsilon_{u,00}}\right)^2 (f_{u,00}^{+,00})^2 = \frac{1}{4}\left(1 - \frac{E_{u,00}}{\epsilon_{u,00}}\right)^2 \frac{\epsilon_{u,00}}{E_{u,00}}$$

and perform the momentum integration, which yields

$$2\pi\lambda \int du\, [\epsilon_{u,00} - E_{u,00}]\,|v_{u,00}^{00}|^2 = \lambda\left(-\frac{\sqrt{2}}{3\pi}\kappa^{3/2} + \frac{\sqrt{2}\ln(4/3)}{\pi}\kappa^{5/2}\right).$$

The first term will provide the one-dimensional beyond-mean-field correction, whereas the second is part of the leading correction. In contrast to the situation with periodic boundary conditions, the leading correction to the one-dimensional behavior is not of order $\kappa^3$, but the coupling of the condensate mode to higher harmonic oscillator modes leads to a three-dimensional behavior of the form $\kappa^{5/2}$. Terms of order $\kappa^3$ appear in our calculations as well but they are not the dominant correction anymore. In the following we will see, that a consistent expansion up to $\kappa^3$ would require to calculate the Bogoliubov functions within second order perturbation theory, which makes a consistent expansion very cumbersome.

We will now continue with the second term of Eq. (12). As $(v,w) \neq (\alpha,\beta) = (0,0)$, we need to calculate the perturbative correction to the Bogoliubov function $f_{u,vw}^{\pm,00}$. In our calculation we determine $f_{u,vw}^{+,00}$ within first order, take care of the normalization in the relevant order and obtain

$$|v_{u,vw}^{00}|^2 = \frac{1}{4}\left(1 - \frac{E_{u,00}}{\epsilon_{u,vw}}\right)^2 (f_{u,vw}^{+,00})^2 = \kappa^2 \frac{\epsilon_{u,00}}{E_{u,00}}\left(1 - \frac{E_{u,00}}{\epsilon_{u,vw}}\right)^2 \left(\frac{\epsilon_{u,vw}\eta_{v0}\eta_{w0}}{\epsilon_{u,00}^2 - \epsilon_{u,vw}^2}\right)^2.$$

Although we calculated the Bogoliubov spectrum $E_{u,00}$ up to second order, it is sufficient to use only the lowest order order, as $v_{u,vw}^{00}$ itself contains orders of $\kappa^2$ and higher. Again, we can perform the momentum integration and expand the result in orders of $\kappa$,

$$2\pi\lambda \int du \sum_{v,w}{}' [\epsilon_{u,vw} - \epsilon_{u,00}]\,|v_{u,vw}^{00}|^2 = \lambda \sum_{vw}{}' \eta_{v0}^2 \eta_{w0}^2 \left(\frac{2\pi^2}{\sqrt{v+w}}\kappa^2 - \frac{16\sqrt{2}\pi}{v+w}\kappa^{5/2} + O(\kappa^3)\right). \tag{14}$$

The first term of order $\kappa^2$ carries the $\kappa$-dependence of the amplitude $|v_{u,vw}^{00}|^2$. Hence, evaluating the Bogoliubov functions up to second order immediately yields additional contributions of order $\kappa^3$. In the following we will waive to include those. The second term is another contribution to the energy of order $\kappa^{5/2}$ with the same double sum as in Eq. (13).

The procedure for the third term of Eq. (12) is very similar to the previous one. It is sufficient to calculate the Bogoliubov spectrum $E_{u,\alpha\beta}$ and the amplitudes $v_{u,00}^{\alpha\beta}$ within first order,

$$E_{u,\alpha\beta}^2 = \epsilon_{u,\alpha\beta}^2 + 2\kappa\epsilon_{u,\alpha\beta}\eta_{\alpha\alpha}\eta_{\beta\beta} \quad \text{and} \quad |v_{u,00}^{\alpha\beta}|^2 = \kappa^2 \frac{\epsilon_{u,\alpha\beta}}{E_{u,\alpha\beta}}\left(1 - \frac{E_{u,\alpha\beta}}{\epsilon_{u,00}}\right)^2 \left(\frac{\epsilon_{u,00}\eta_{\alpha 0}\eta_{\beta 0}}{\epsilon_{u,\alpha\beta}^2 - \epsilon_{u,00}^2}\right)^2.$$

The expansion for small values of $\kappa \ll 1$ and the momentum integration yields another contribution of order $\kappa^2$,

$$2\pi\lambda \int du \sum_{\alpha,\beta}{}' [\epsilon_{u,00} - E_{u,\alpha\beta}] |v_{u,00}^{\alpha\beta}|^2 = -\lambda \sum_{\alpha,\beta}{}' 2\pi^2 \frac{\eta_{\alpha 0}^2 \eta_{\beta 0}^2}{\sqrt{\alpha+\beta}} \kappa^2 + O(\kappa^3), \tag{}$$

which exactly cancels the $\kappa^2$ contribution of Eq. (14).

For the last term in Eq. (12), even the lowest order in the Bogoliubov spectrum $E_{u,\alpha\beta}$ and the amplitudes

$$|v_{u,vw}^{\alpha\beta}|^2 = \frac{1}{4}\left(1 - \frac{E_{u,\alpha\beta}}{\epsilon_{u,vw}}\right)^2 \frac{\epsilon_{u,vw}}{E_{u,\alpha\beta}} \delta_{v\alpha}\delta_{w\beta} \tag{15}$$

immediately leads to contributions of order $\kappa^3$.

In conclusion, we arrive at a differential equation for $\kappa \ll 1$,

$$E_{1D}^h - \frac{1}{2}\mu N = E_{1D}^h - \frac{\kappa}{2}\frac{dE_{1D}^h}{d\kappa} = \lambda E_0^h \left(-\frac{\sqrt{2}}{3\pi}\kappa^{3/2} - \frac{\sqrt{2}\ln(4/3)}{\pi}\kappa^{5/2} + O(\kappa^3)\right) \tag{16}$$

and its solution reads

$$\frac{E_{1D}^h}{E_0^h} = \lambda \frac{C_{1D}^h}{\sqrt{2}}\kappa^2 - \lambda \frac{4\sqrt{2}}{3\pi}\kappa^{3/2} + \lambda \frac{4\sqrt{2}\ln(4/3)}{\pi}\kappa^{5/2} + O(\kappa^3). \tag{17}$$

Note that the first term of order $\kappa^2$ is not determined by the differential equation but enters as a constraint, as the crossover has to include the leading correction to the confinement-induced resonance, which became evident in the study for periodic boundary conditions. For a harmonic confinement, we confirm this by a full numerical evaluation of the ground state energy in the crossover form 3D to 1D. This requires the full numerical evaluation of the Bogoliubov excitation spectrum $E_{u,\alpha\beta}$ and the determination of the factors $|v_{u,vw}^{\alpha\beta}|^2$. Inserting the result in Eq. (11) and performing the summation and integration numerically allows us to solve the differential equation and obtain the ground state energy $E_{1D}^h$ in the full crossover. We can fix the integration constant at large densities ($\kappa \gg 1$) by requiring the correct mean-field term from the local-density approximation. Then, we can derive the behavior in the one-dimensional regime $\kappa \ll 1$. From our numerical calculations, we obtain the leading corrections in Eq. (14) in the main text, and recover the analytical expressions for $C_{1D}^h$ with an error of 1%. Remarkably, from the 3D result in local-density approximation, we recover from the crossover the expected LHY correction, as well as the correction stemming from the regularization of the 1D scattering length due to the transverse confinement.

Finally, we want to comment on the situation in the 2D geometry with harmonic confinement. The analysis of the ground state energy in the two-dimensional regime is carried out analogously to the 1D scenario. We find, that the leading correction to the beyond-mean-field energy is of order $\kappa^3 \ln(\kappa)$. This is problematic as a consistent expansion up to order $\kappa^3$ would require to determine the Bogoliubov functions $f_{\mathbf{k},w}^{\pm,\alpha}$ within second order perturbation theory, as we have already seen in the one-dimensional case.



\end{document}

%% file: paper-LHY-crossover.bbl
%